# A Theoretical Prediction on the Intrinsic Half-Metallicity in the Surface-Oxygen-Passivated $Cr_2N$ MXene

Guo Wang

Department of Chemistry, Capital Normal University, Beijing 100048, China

Email: wangguo@mail.cnu.edu.cn

ABSTRACT: Two-dimensional $Cr_2N$ MXene as well as the surface-passivated $Cr_2NF_2$, $Cr_2N(OH)_2$ and $Cr_2NO_2$ are investigated by using density functional theory. The $Cr_2N$ is an anti-ferromagnetic metal. The F atom or OH group-passivation does not change the anti-ferromagnetic characteristics. However, $Cr_2NO_2$ has a ferromagnetic ground state, which is a half-metal. The half-metallicity of $Cr_2NO_2$ is still robust when bias is applied to the nanometer sized device. The half-metallicity is intrinsic and does not require atomically clean surfaces. Therefore the stable surface-oxygen-passivated MXene is a good candidate for spintronics.

# INTRODUCTION

The MAX phases are tens of ternary carbides and nitrides with the formula $M_{n+1}AX_n$ (n=1-3), in which A is an A-group element and M is a transition metal element. They combine several outstanding properties of metals and ceramics, such as readily machinable (with hack-saw), high damage tolerance, high electrical conductivity, as well as high-temperature oxidation resistance, making them potential candidates for many applications.[1] The MAX phases are hexagonal layered materials, which have two-dimensional characteristics. Usually, the properties of two-dimensional monolayer, such as graphene, are drastically different from those of bulk materials.[2] In the MAX phases, the A-group elements are weakly bonded and reactive. The A-group can be selectively etched and then graphene-like MXenes can be exfoliated.[3] Because of the unique structures, MXenes have potential applications in sensors, catalysts and electrochemical energy storage.[4]

Recent theoretical investigations indicate that several MXenes,[4] such as $Ti_3C_2$,[5-7] $Ti_3N_2$,[5] $Ti_3CN$,[7] $Ti_2C$, $Ti_2N$,[8,9] $Ta_3C_2$,[10] $Cr_2C$ and $Cr_2N$,[11] as well as MXene nanoribbons[12] have magnetic ground states. They have interesting magnetism-related properties, such as half-metallicity of $Cr_2C$[13] and near half-metallicity of $Ti_2C$ and $Ti_2N$.[9] For $Ti_2C$, the structure becomes a true half-metal with a minority-spin gap of 0.04 eV under a biaxial strain.[9] Moreover, $Cr_2C$ has intrinsic half-metallicity with a large minority-spin gap of 2.85 eV,[13] which does not need external conditions, such as strong electric field[14] or doping[15]. Therefore, $Cr_2C$ becomes a promising two-dimensional candidate for spintronic devices. It is noted that the half-metallicity requires atomically clean surfaces. When the surfaces are passivated, the half-metallicity of $Cr_2C$ is suppressed.[13] For the above titanium-containing MXenes, the magnetism also disappears when the bare surfaces are passivated.[4-8] During etching process, for example with hydrofluoric acid,[3] it is difficult to maintain such clean surfaces. This limits their application in spintronics. In the present work, another MXene $Cr_2N$, which has a different number of electrons in a unit cell with

$Cr_2C$, is investigated by using density functional theory. It is indicated that the bare $Cr_2N$ is not half-metallic. On the contrary, half-metallicity appears when the surfaces are passivated by oxygen atoms.

## MODELS AND COMPUTATIONAL DETAILS

The $Cr_2N$ MXene is assumed to be exfoliated from parent $Cr_2GaN$. During the supposed treatment with hydrofluoric acid, the Ga atoms should be replaced by F or OH. On the other hand, $Cr_2GaN$ is also observed to extrude elemental Ga at room temperature. Gases, such as $O_2$, are assumed to replace the Ga in the basal plane.[16] Thus the bare $Cr_2N$ and the surface-passivated $Cr_2NT_2$ (T=F, OH or O) are investigated.

As shown in Figure 1a,b, two-dimensional bare $Cr_2N$ has a hexagonal layered structure and the N atom layer is sandwiched between the two layers of the Cr atoms. The Cr atoms in different layers are marked with different colors. Both ferromagnetic (FM) and anti-ferromagnetic (AFM) configurations are considered in the calculations, which are shown in Figure 1c-h. In the FM, AFM2 and AFM4 configurations, the spins in the different Cr layers are parallel to each other. On the contrary, the spins in the different Cr layers are anti-parallel in the AFM1, AFM3 and AFM5 configurations. As denoted by tetragons in Figure 1c-h, only the primitive cell is needed for FM and AFM1 configurations. A 2×1 unit cell is constructed for calculating AFM2 and AFM3 configurations, while a 2×2 unit cell is used to calculate AFM4 and AFM5 configurations.

The geometrical optimization and the electronic properties calculations are performed by the VASP program.[17] The PBE density functional[18] and the projector-augmented wave basis with an energy cutoff of 400 eV are used. In order to describe the strongly correlated 3d electrons of Cr atoms, the GGA+U method is used. In this framework, the difference between the onsite coulomb (U) and exchange (J) parameters is set to 3 eV (U-J). This value is successfully applied to similar systems, such as $CrO_2$[19] and $CrC_2$[13]. A 41×41×1 Monkhorst-Pack k-point sampling is used in the first Brillouin zone of the hexagonal unit cell. The vacuum layer is around 15 Å in order to avoid possible interaction between images. The HSE06 hybrid density functional,[20] which can accurately describe the properties of solids,[21] is also used to confirm the energies and electronic properties obtained by the GGA+U method. In order to further verify the stability, the phonon dispersion is obtained through density functional perturbation theory with the aid of Phonopy code,[22] which is based on a 4×4 supercell.

Besides the infinitely long structures, electronic devices with size of nanometer are also investigated based on the non-equilibrium Green's function method. The current-bias characteristics is obtained through the Landauer-Büttiker formula[23] by using the OPENMX program[24]

$$I = \frac{e}{h} \int_{-\infty}^{+\infty} T(E,V_b)[f_L(E) - f_R(E)]dE$$

in which $T(E, V_b)$ is a transmission coefficient at energy $E$ and bias $V_b$, $f_L(E)$ and $f_R(E)$ are Fermi-Dirac distribution functions for left and right electrodes. Two-dimensional rectangular unit cell is constructed, which is composed of two primitive cells. The left and right electrodes are composed of two unit cells while the scattering region is composed of three unit cells. Norm-conserving pseudopotentials[25] and variationally optimized pseudo-atomic localized basis functions[26, 27] are used in this program. Each s, p or d (for Cr atom) orbital is described by a primitive orbital. Default convergence criterions are used in the calculation. The self-consistent field iteration stops when the absolute energy deviation between the current and the previous step is less than $10^{-6}$ Hartree. The energy cutoff is 150 Ry and 121 k-points are used along the transport direction. The GGA+U method (the PBE density functional with U-J=3 eV) is also available in this program. For consistency, the results are all obtained by the GGA+U method unless explicitly stated.

## RESULTS AND DISCUSSION

For the FM configuration, the optimized bond length between the Cr and N atoms is 2.13 Å. In its band structures shown in Figure 2a, there are bands going across the Fermi level for both spins. It is a metal, but not a half-metal. The number of electrons in a N atom is bigger than that in a C atom, so the band structures are different from those of the half-metallic $Cr_2C$.[13] The unbalanced band structures for both spins imply magnetic properties. The total magnetic moment is 8.61 $\mu_B$. Detailed analysis indicates that the magnetism is mainly located on Cr atoms. Each Cr atom has a magnetic moment of 4.45 $\mu_B$, while that on a N atom is -0.30 $\mu_B$. These values are not integers, which are different from those of $Cr_2C$.[13] In order to elucidate the fractional magnetic moment, Bader charge analysis[28-30] is performed. The net charge on a N atom is -1.50 |e|. Unlike the C atom in $Cr_2C$, the N atom with fractional charges possesses non-zero magnetic moment. The Cr atom with a formal fractional charge +0.75 also has a fractional magnetic moment.

The relative energies of the optimized FM, AFM1, AFM2, AFM3, AFM4 and AFM5 structures are 203, 301, 175, 0, 1394 and 1568 meV, respectively. This indicates that $Cr_2N$ has an AFM3 ground state. There are two types of bond lengths (2.07 and 2.15 Å) in the optimized AFM3 structure. The symmetry breaking and geometrical distortion introduce stronger bonds, especially for the Cr and N atoms with distance of 2.07 Å, which introduces additional stability. It should be noted that the initial hypothetical AFM4 and AFM5 configurations shown in Figure 1g,h have never been obtained by various initial structures. This indicates that the structures with so many anti-ferromagnetic couplings are not stable. The structures either converges to the most stable AFM3 configurations, or to the nominal AFM4 and AFM5 configurations shown in Figure 1i,j. In the nominal AFM4 and AFM5 configurations, the spins are localized on only a few Cr atoms in the distorted structures. The absence of the magnetism on the other Cr atoms makes the two nominal configurations very unstable. Only AFM1, AFM2 and AFM3 configurations are normal AFM states.

All the above AFM $Cr_2N$ structures have anti-ferromagnetic band structures. Only those of the most stable AFM3 structures are shown in Figure 3a. Since the band structures for majority and minority spins are degenerate, only those for the majority spin are shown. In the band structures, two bands go across the Fermi level, indicating metallic characteristics. Many AFM structures are non-metallic. Two-dimensional $Cr_2N$ is an AFM metal, which should be useful in spintronics. However, the bare surfaces can be passivated during etching process. Its electronic properties may change after surface-passivation.

The hybrid density functional HSE06 is also used to confirm the results based on the GGA+U method. In the FM configuration, the bond length is 2.10 Å, close to the GGA+U result (2.13 Å). The total magnetic moment is 8.59 $\mu_B$, also close to the GGA+U result (8.61 $\mu_B$). The relative energies per primitive cell of the optimized FM, as well as the normal AFM1, AFM2 and AFM3 structures are 295, 336, 130 and 0 meV. AFM3 configuration is also the ground state. The optimized AFM3 structure also has two types of bond lengths (2.04 and 2.11 Å). The HSE06 density functional indicates that the two-dimensional $Cr_2N$ is an AFM metal. The results obtained by the HSE06 density functional are consistent with those obtained by the GGA+U method.

In the surface-passivated structures, the functional atoms or groups are above the hollow site of the N atoms and pointed to the Cr atoms in the opposite layer.[8, 11] The relative energies per primitive cell of the optimized FM, AFM1, AFM2, AFM3, AFM4 and AFM5 structures are 58, 543, 200, 0, 3890 and 1538 meV for $Cr_2NF_2$, 38, 260, 149, 0, 1427 and 1391 meV for $Cr_2N(OH)_2$, 0, 80, 220, 264, 1782 and 2609 meV for $Cr_2NO_2$. Similar to the situation of $Cr_2N$, the nominal AFM4 and AFM5 configurations are very unstable for the surface-passivated structures. From bare $Cr_2N$ to the mono-valent functional group-passivated $Cr_2NF_2$ and $Cr_2N(OH)_2$, the AFM3 configurations are always the ground states. However, the divalent oxygen atom-passivated $Cr_2NO_2$ has a FM ground state.

The band structures of $Cr_2NF_2$ and $Cr_2N(OH)_2$ with stable AFM3 configurations shown in Figure 3b,c are quite similar to those of $Cr_2N$. Although the AFM3 configuration is not the ground state of $Cr_2NO_2$, the band structures are also shown in Figure 3d for comparison. Like the other AFM3 structures, there are two bands going across the Fermi level in the band structures of $Cr_2NO_2$. The four configurations are all AFM metals.

As shown in Figure 2d, the band structures of $Cr_2NO_2$ with FM ground state indicate that it is a half-metal. It is metallic for majority spin and semiconducting for minority spin. Actually, the band structures of $Cr_2NF_2$ and $Cr_2N(OH)_2$ with FM configuration (shown in Figure 2b,c) also have this half-metallic characteristics. As shown in Figure 2a, the bare $Cr_2N$ with FM configuration is metallic for both spins. After the surface-passivations, the total magnetic moments in the FM configurations drops from 8.61 to 7.00 $\mu_B$ both for $Cr_2NF_2$ and $Cr_2N(OH)_2$. The magnetic moment on each Cr atom is 3.59 $\mu_B$. The magnetic moment on the N atom is -0.23 $\mu_B$, while it is 0.02 or 0.01 $\mu_B$ on each F or O atom. Compared with $Cr_2N$, the magnetic moment on the Cr atom reduces by about 1 $\mu_B$ (from 4.45 to 3.59 $\mu_B$). This is attributed to the surface-passivation by mono-valent functional atoms or groups. For the divalent oxygen atom-passivated $Cr_2NO_2$, the total magnetic moment reduces further to 5.00 $\mu_B$. The magnetic moment on each Cr atom is 2.76 $\mu_B$, while that on each N or O atom is -0.18 or -0.17 $\mu_B$. With the functional atoms or groups, the electrons with minority spin at the Fermi level are suppressed in the FM configuration of $Cr_2N$. From $Cr_2N$ to $Cr_2NF_2$, $Cr_2N(OH)_2$ and then to $Cr_2NO_2$, the total magnetic moments and those on each Cr atom drop. The relative energies of the FM configurations with respect to the corresponding AFM3 configurations decrease from 203 to 58 and 38 meV for $Cr_2N$, $Cr_2NF_2$ and $Cr_2N(OH)_2$. Finally, the FM configuration becomes the ground state of $Cr_2NO_2$. A possible route to obtain $Cr_2NO_2$ is extruding elemental Ga from $Cr_2GaN$ at room temperature.[16] In the etching technology, $Cr_2NO_2$ is assumed to be obtained by a post treatment of the resulting $Cr_2N(OH)_2$. The energy change of the reaction $Cr_2N(OH)_2 + 1/2\ O_2 \rightarrow Cr_2NO_2 + H_2O$ is -1.7 eV, indicating that this process is energetically favorable.

For the half-metallic $Cr_2NO_2$, the HSE06 density functional is also used to confirm the result obtained by the GGA+U method. After geometrical optimization, the band structures shown in Figure 2e are quite similar to those in Figure 2d. The half-metallicity of $Cr_2NO_2$ maintains. There are still two bands going across the Fermi level only for majority spin. The half-metallicity is intrinsic and it does not need external strong electric field or doping.[14, 15] And more importantly, $Cr_2NO_2$ is surface-passivated and is more likely to be realized than the bare structure. Therefore, the surface-oxygen-passivated $Cr_2N$ MXene should be a promising candidate for spintronics. The indirect band gap of $Cr_2NO_2$ for minority spin is 2.79 or 3.88 eV under the GGA+U or HSE06 method. The large band gap implies sustainable half-metallicity during device operation.

The size of an electronic device becomes progressively smaller with increasing time. It is interesting to see whether the half-metallicity is sustainable when bias is applied to nanometer sized devices. The transmission spectrum at zero bias of the $Cr_2NO_2$ device is shown in Figure 4a. In such a narrow device, the transmission for minority spin is still absent and the half-metallicity maintains. As shown in Figure 4b, the current for majority spin or the total current increases with the increasing bias until the bias reaches 0.5 V. Then the current is saturated when the bias increases further to 1 V. The spin polarization is always 100% at this bias range. The half-metallicity of the $Cr_2NO_2$ is still robust when a wide range of bias is applied to the device. The current at 0.5 V is 19.7 μA, corresponding to a high conductance of 39.4 μS. The nanometer sized $Cr_2NO_2$ device is highly conductive and has an 100% spin polarization. It should be a good candidate for spintronic devices.

Since $Cr_2NO_2$ has appealing properties, its phonon dispersion is calculated. As shown in Figure 5, no non-trivial imaginary frequency is found. The imaginary frequencies (no more than 3 cm$^{-1}$) near the Γ points are attributed to the three translational modes. Because of the numerical method adopted by density functional theory, the frequencies of the translational modes are not exactly zero. The phonon dispersion verifies the stability of $Cr_2NO_2$. Ab initio molecular dynamics simulations are also performed. The VASP program currently can only perform constant volume molecular dynamics. A micro canonical ensemble is used for the four stable configurations. A 2×2 unit cell is used and the simulation temperature is set to 300 K. The step size is set to 1 fs. During the 5000 simulation steps, no essential structural deformation is found except the orientation of the hydrogen atoms in $Cr_2N(OH)_2$. In Figure 6a-d, the energy fluctuation ranges are 193, 264, 412 and 368 meV per primitive cell for $Cr_2N$, $Cr_2NF_2$, $Cr_2N(OH)_2$ and $Cr_2NO_2$, respectively. These values are higher than the energy differences between different corresponding configurations. However, the transition between different configurations does not occur during the molecular dynamics simulations. This confirms the stabilities of $Cr_2NO_2$ with FM configuration, as well as $Cr_2N$, $Cr_2NF_2$ and $Cr_2N(OH)_2$ with AFM3 configurations. The highest energy fluctuation range for $Cr_2N(OH)_2$ should be attributed to the random orientation of the hydrogen atoms in the molecular dynamics simulation. Although the AFM3 ground state of $Cr_2N$ is generally unchanged, the total magnetic moment changes slightly. As shown in Figure 6e, the total magnetic moment oscillates up and down. The amplitude is in a rough range of 0.2-0.6 $\mu_B$. After the surface-passivation by mono-valent F atoms or OH groups, the oscillation becomes much weaker. In Figure 6f, the total magnetic moment of $Cr_2NF_2$ is always zero except a very small deviation of -0.01 $\mu_B$ around 3919 fs. The magnetic configuration is much more stable than that of the bare $Cr_2N$. For $Cr_2N(OH)_2$, the largest deviation as shown in Figure 6g is -0.11 $\mu_B$ and this value is much smaller than that of the bare $Cr_2N$. The slightly stronger oscillation compared to $Cr_2NF_2$ should be also attributed to the random orientation of the hydrogen atoms. For the divalent oxygen atom-passivated $Cr_2NO_2$, the total magnetic moment as shown in Figure 6h is unchanged. The molecular dynamics simulations also indicate that the surface-passivation, especially the surface oxygen atom-passivation stabilizes the total magnetic configurations of the $Cr_2N$ MXenes.

## CONCLUSIONS

Two-dimensional $Cr_2N$ MXene as well as the surface-passivated MXenes are investigated by using density functional theory. The bare $Cr_2N$ MXene is an AFM metal. The surface-passivated $Cr_2NF_2$ and $Cr_2N(OH)_2$ are also AFM metals, while $Cr_2NO_2$ is a half-metal. The number of electrons in $Cr_2N$ is different from that in $Cr_2C$. Unlike $Cr_2C$ that is a half-metal, the FM configuration of the $Cr_2N$ is metallic for both spins. After surface-passivation, the electrons with minority spin at the Fermi level are suppressed and the MXenes with FM configurations have half-metallic band structures. From the bare $Cr_2N$ to the mono-valent atom or functional group-passivated $Cr_2NF_2$ and $Cr_2N(OH)_2$, and then to the divalent oxygen atom-passivated $Cr_2NO_2$, the FM configuration becomes more and more stable. Finally, $Cr_2NO_2$ has a FM ground state. The half-metallicity of the $Cr_2NO_2$ MXene is intrinsic, which does not require external electric field or doping. Because it has a large band gap of 2.79 eV for minority spin, the half-metallicity is even robust when bias is applied to the nanometer sized $Cr_2NO_2$ device. Unlike the $Cr_2C$, the half-metallicity here does not require atomically clean surfaces. The $Cr_2NO_2$ MXene is a surface-passivated stable structure. Therefore, the surface-oxygen-passivated $Cr_2N$ MXene should be a promising candidate for spintronics.

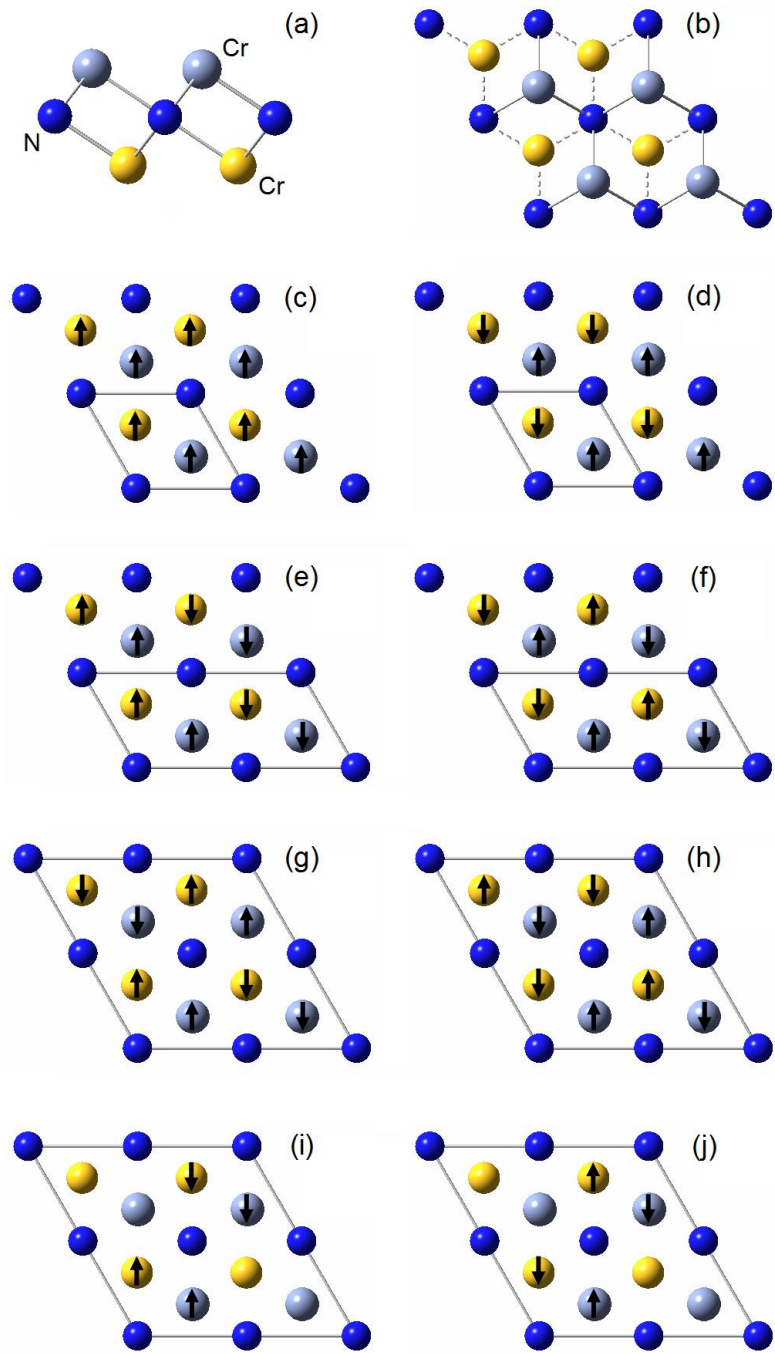

**Figure 1.** (a) Side and (b) top view of two-dimensional Cr$_2$N, (c) FM, (d) AFM1, (e) AFM2 and (f) AFM3 configurations, (g) initial guess of AFM4 and (h) AFM5 configurations, (i) AFM4 and (j) AFM5 configurations for the optimized structures. The adopted unit cells are denoted by tetragons in (c)-(j). The Cr atoms in different layers are labeled with different colors.

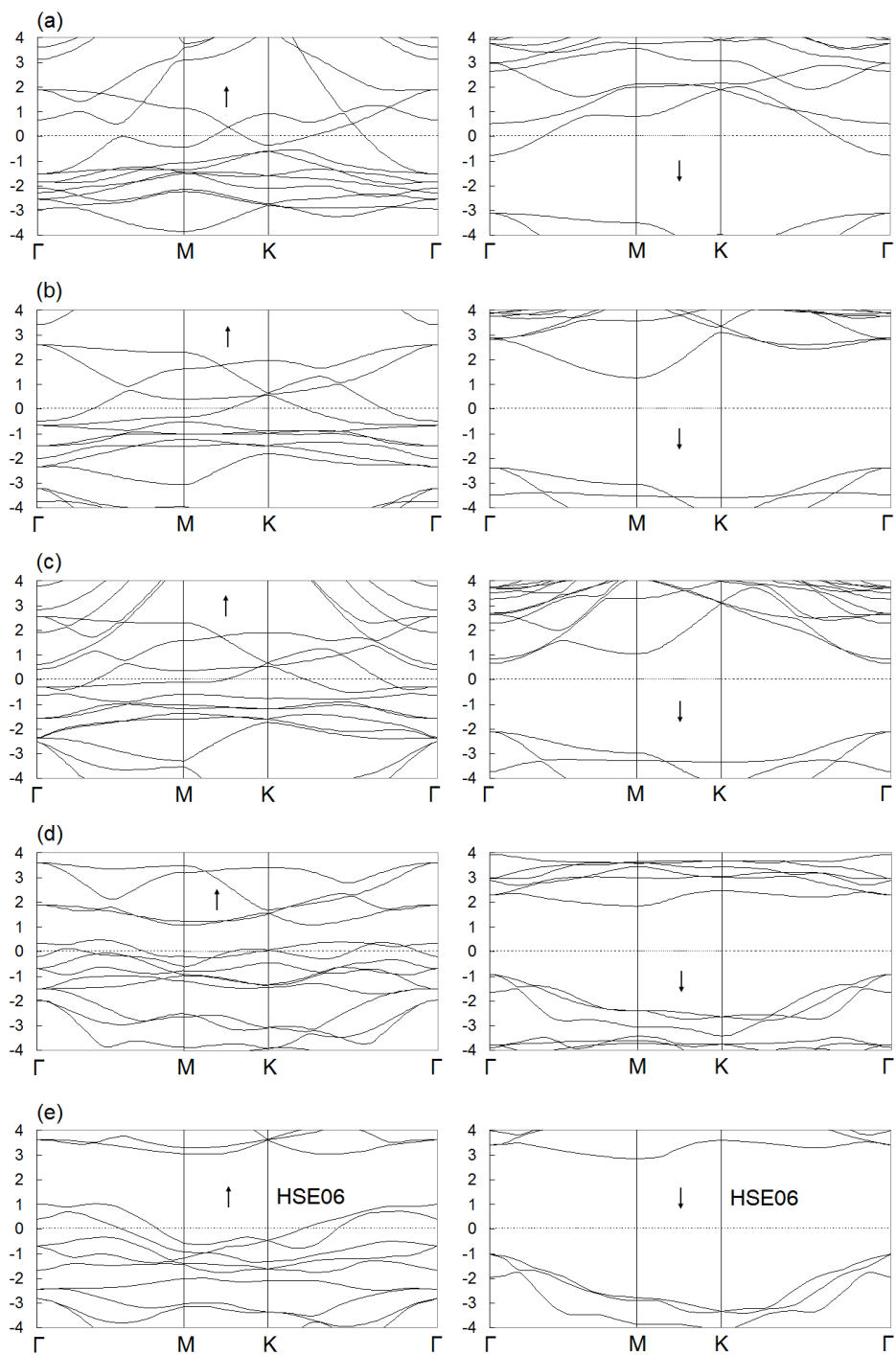

**Figure 2.** Band structures of (a) $Cr_2N$, (b) $Cr_2NF_2$, (c) $Cr_2N(OH)_2$ and (d) $Cr_2NO_2$ with FM configurations, (e) band structures of $Cr_2NO_2$ with FM configurations obtained by the HSE06 density functional. Vertical axis: energy in eV, horizontal axis: reciprocal lattice vector.

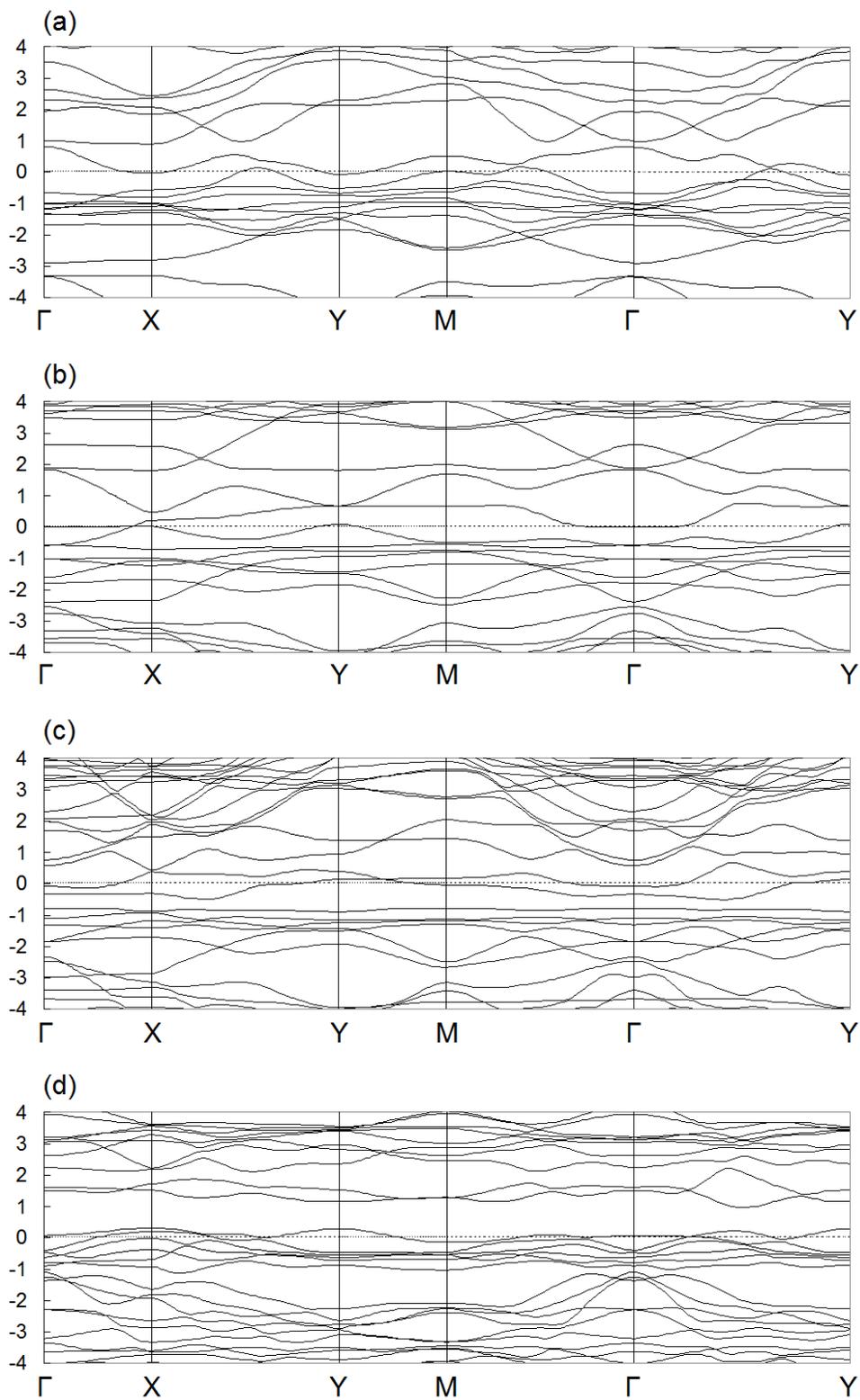

**Figure 3.** Band structures of (a) Cr$_2$N, (b) Cr$_2$NF$_2$, (c) Cr$_2$N(OH)$_2$ and (d) Cr$_2$NO$_2$ with AFM3 configurations. Vertical axis: energy in eV, horizontal axis: reciprocal lattice vector.

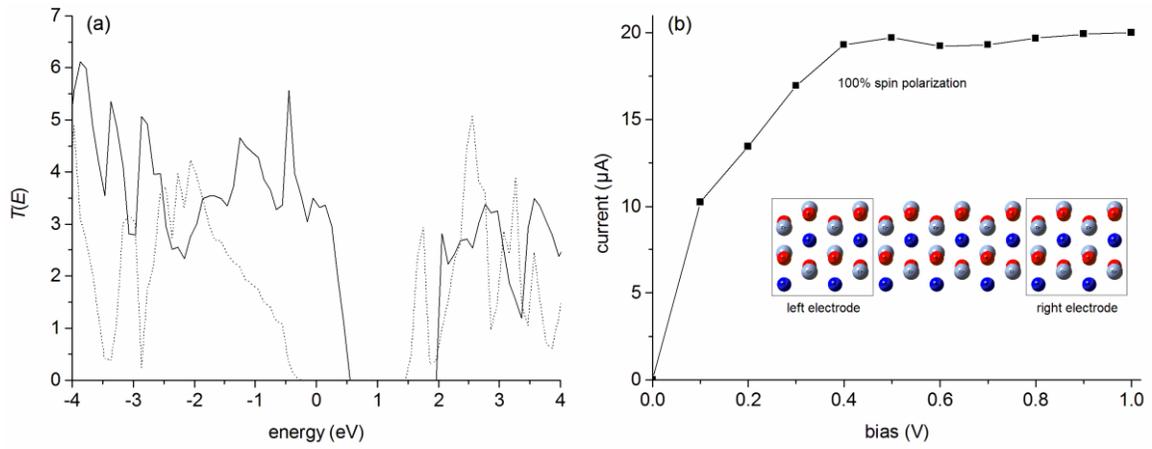

**Figure 4.** (a) Transmission spectrum and (b) current-bias curve of the $Cr_2NO_2$ device, structure of the device is in the inset.

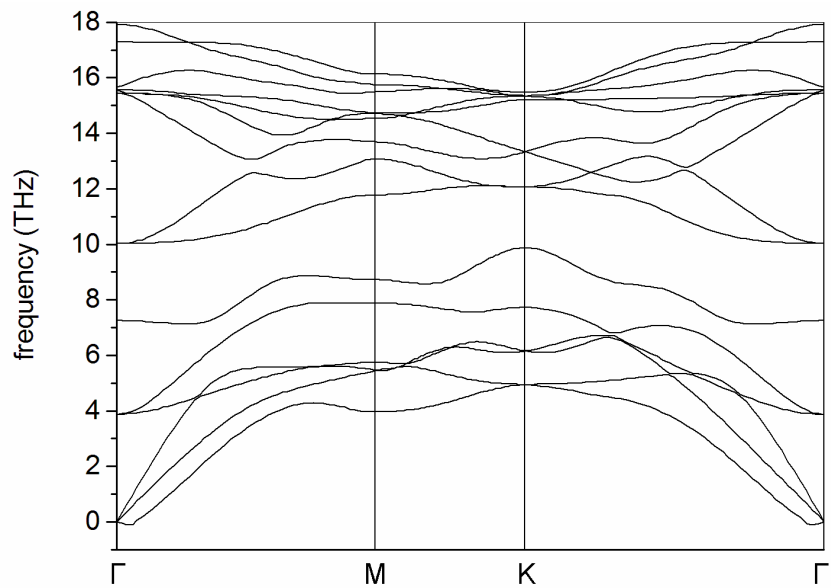

**Figure 5.** (a) Phonon dispersion of $Cr_2NO_2$ with FM configuration.

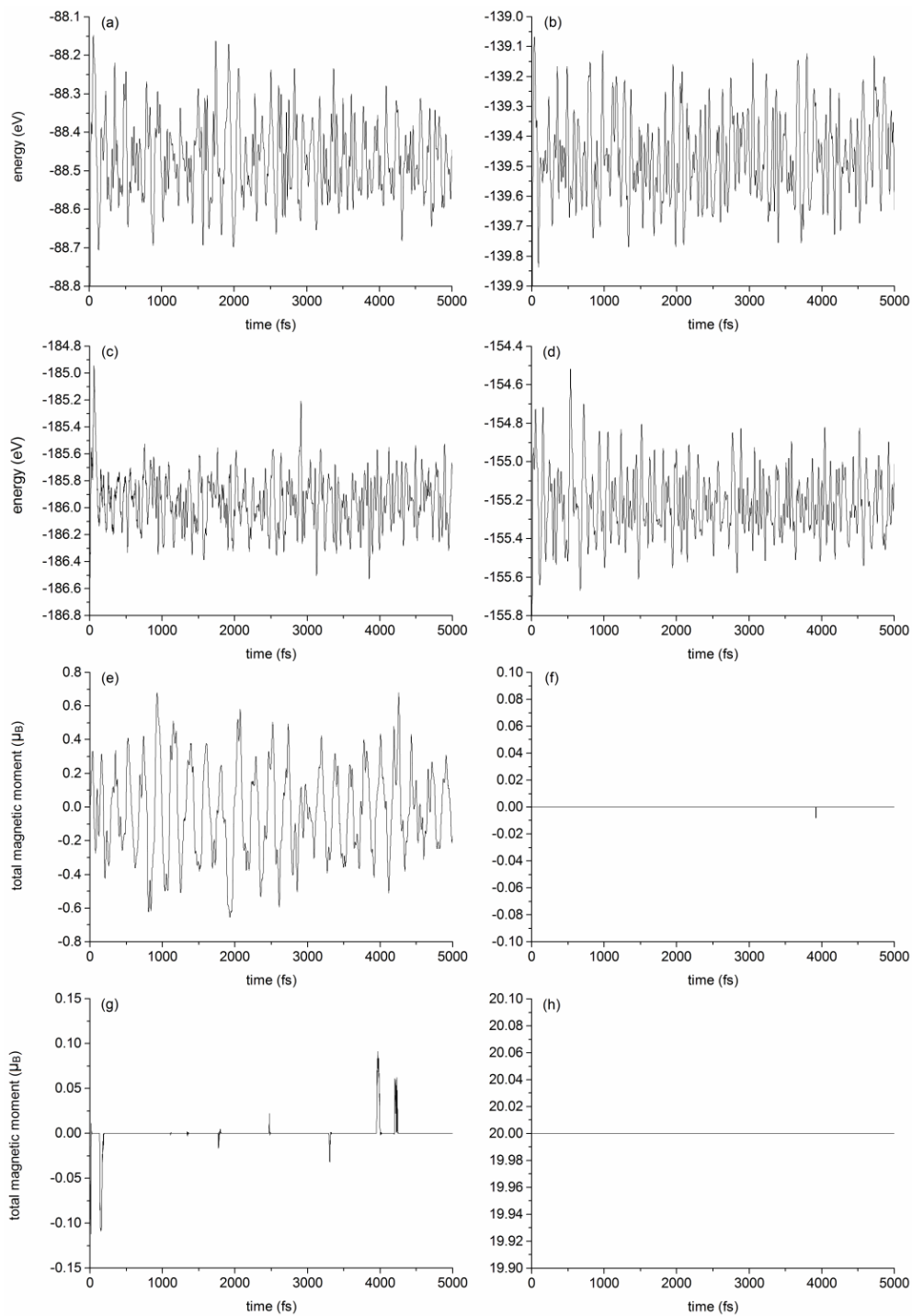

**Figure 6.** Energy fluctuation of (a) $Cr_2N$ (b) $Cr_2NF_2$, (c) $Cr_2N(OH)_2$, (d) $Cr_2NO_2$ and total magnetic moments of (e) $Cr_2N$ (f) $Cr_2NF_2$, (g) $Cr_2N(OH)_2$, (h) $Cr_2NO_2$ during molecular dynamic.